\documentstyle[manuscript,epsfig,aps]{revtex} 
\tightenlines 
 
\begin{document} 
\title{Correlation effects in Co/Cu and Fe/Cr magnetic multilayers} 
\author{L. Chioncel$^1$ and A. I. Lichtenstein$^{1,2}$} 
\address{ $^1$ University of Nijmegen, NL-6525 ED Nijmegen, The Netherlands \\
$^2$ Institut f\"ur Theoretische Physik, Universitaet Hamburg, 20355 Hamburg , Deutchland} 
\date{27 May 2003} 
\maketitle 
 
\begin{abstract} 
The electronic structure of Co/Cu(001) and Fe/Cr(001) magnetic multilayers
has been investigated within the local density approximation combined with
dynamical mean field theory. Our calculation shows enhanced density of 
states at the Fermi level, suggesting that electronic correlations might
play an important role in the transport properties of multilayers.
\end{abstract} 
 
 
\section{Introduction} 

Magnetic multilayers (MML) heterostructures of alternating ferromagnetic layers 
and non-magnetic spacers have  attracted attention in the last decades  because
of their implications both for fundamental research and technological applications. 
The most remarkable  property of these systems is  the giant  magnetoresistance
(GMR) measured for a parallel/antiparallel configuration of the magnetic moments
belonging to different layers in the presence of an external magnetic filed \cite{GMR}.
The GMR effect is related to the spin dependent scattering but the detailed
mechanism is still subject of intense investigations. 

One of the first attempts to explain the GMR effect was based on the ballistic
approach  \cite{Schep95}. Soon it became obvious that the band structure 
description is very important for the realistic description of GMR. A 
considerable number of 
attempts have been made to include the electronic structure in diffusive
regime \cite{Butler93,Nesbet94}. In order to consider the spin dependent scattering 
due to the interface, these calculations were based on the coherent potential
approximation. However, all these calculations overestimate the experimental values
for the GMR effect. On the other hand band structure calculations allows the evaluation
of the Fermi velocities which can be used to estimate the spin dependent 
relaxation-times \cite{Zahn95}. The 
semiclassical approach used in the calculation of the conductivity overestimate
also, the experimental GMR values \cite{Tsymbal97}. 
One of the most advanced technique in the {\it ab initio} theories of electric
transport in solid systems, is offered by the Kubo-Greenwood formalism. A 
comprehensive overview of this technique applied for systems with reduced 
dimension is discussed by Weinberger \cite{Weinberger2003}. 

In this paper, we demonstrate that the correlations effects might play an
important role in the realistic description of transport properties. Our
approach is motivated by the fact that the  $3d$ transition metal elements,
components of the magnetic multilayers, show significant correlation 
effects. In general, spin-polarized LDA band structure calculation gives an adequate
description of the ferromagnetic ground state for the most of metals but, on the
other hand, there are obvious evidences of essentially many-body features in
photoemission spectra of Fe \cite{Katsnelson99}, Co \cite{Co}, and Ni
\cite{Lichtenstein01},\cite{Antropov}. The 6 eV satellite in Ni, broadening of the ARPES 
features due to quasiparticle damping, narrowing of the $d$-band, essential 
change of spin polarization near the Fermi level are some examples. It is obvious
that these effects can be equally important also for magnetic multilayers and
other heterostructures containing transition metals.

\section{Numerical details}

We focus on the band structure of the perfect (001) fcc $Co_6/Cu_5/Co_5$ 
and (001) bcc Fe$_3$/Cr$_5$ supercells. It was found experimentally that values of GMR are
$220 \%$ in Fe/Cr multilayers \cite{Schad95} and $120 \% $ in Co/Cu multilayers
\cite{Parkin91}. One of the reasons why the above multilayers are highly magnetorezistve
is that they contain ferromagnetic 3d metals which should have a pronounced spin
 asymmetry in their conductivity due to the presence of the exchange split $d$-bands. 
Perhaps the crucial factors for obtaining high values of GMR are the band matching and the 
lattice matching between the ferromagnetic and non-magnetic metals 
\cite{Tsymbal96,Tsymbal97}. These two conditions
are almost perfectly satisfied in Co/Cu and Fe/Cr multilayers. Thin films of Co grow 
in the fcc structure with the lattice parameter of $3.56 \AA$, which is only $2\%$ 
less than the lattice parameter of $3.61 \%$ in fcc Cu. Both Fe and Cr have the bcc 
structure and their lattice parameters are almost identical: $2.87 \AA$ for Fe and 
$2.77 \AA$ for 
Cr. On the other hand it was experimentally demonstrated that the magnetic multilayers 
grown epitaxially shows an enhancement of the electronic contribution to the low
temperature specific heat \cite{Revaz02} which can be an evidence of correlation effects. 
Indeed, this enhancement cannot be reproduced in the standard electronic band
structure calculations \cite{Kulikov97}, probably because of the neglecting
of many-body effects. It is our aim to check whether the correlation effects considered
in the Dynamical Mean Field Theory (DMFT) \cite{Georges96} approach can lead to an essential 
renormalization of the density of state at the Fermi level $N(E_{F})$. 

We performed calculations based on the LDA+DMFT scheme \cite{DMFT}.
The present realistic LDA+DMFT method \cite{EMTODMFT} is based on 
the so-called Exact Muffin-tin orbitals (EMTO)
method \cite{andersen94,vitos00} within a screened KKR \cite{weinberger90,Taylor83},
frozen core together with the selfconsistent local spin density approximation (LDA).
The correlation effects are treated in the framework of DMFT \cite{Georges96},
with the Spin-Polarized T-matrix plus the Fluctuation Exchange (T-FLEX) approximation
for the quantum impurity solver \cite{Katsnelson01,berlin}.
 
\begin{figure}
\centerline{\psfig{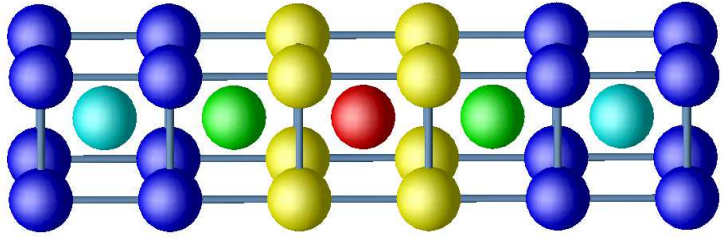}}
\caption{The superlattice strcuture of $Fe_3/Cr_5$ having an equidistant distribution
of atomic layers. $Fe$ atoms are denoted by blue spheres and $Co$ atoms are indicated by green
yellow and red sphere, in the following order: $Fe_1/Fe_2/Fe_1/Cr_1/Cr_2/Cr_3/Cr_2/Cr_1/Fe_1/Fe_2/Fe_1$.
The tetragonal supercell is formed by the first 8 atomic layers (3Fe+5Cr) 
aligned along the $z$ direction.}
\label{fecr_str}
\end{figure}

In the calculations for the $Co/Cu$ system we considered a tetragonal supercell 
formed by 16 layers, each layer containg one atom, with an interlayer distance 
corresponding to the fcc $Cu$ lattice constant $3.61 \AA$.
The supercell structure used for the 
study of the $Fe_3/Cr_5$ system is presented in Fig. \ref{fecr_str}. As can be seen 
the optimization (relaxation) of the atomic layers are neglected, the interlayer distances
correspond to the value of the bulk bcc Fe  $2.88 \AA$. Each atom type is located on 
one layer, magnetically symmetric atoms are represented by the same colored spheres.
$Fe_1$ atoms, or interface atoms, are denoted by blue sphere. The central, $Fe_2$ atoms 
are indicated by a light blue color. In the picture representing 
the structure Fig. \ref{fecr_str}
three extra Fe layers: $Fe_1/Fe_2/Fe_1$ belonging to the next unit cell were introduced. 
Interface $Cr_{1,2}$ layers are indicated by green, respectively 
yellow color spheres, meanwhile
the central $Cr_3$ layer is denoted by red sphere. In the calculations the same atomic
sphere radius was considered for Fe and Cr in the Fe/Cr, and similarly 
the radii for Co and Cu 
atoms were chosen the same for the Co/Cu system.

The $3s, 3p$, and $3d$ states are included for the Co, Cu and Fe valence electrons.
In order to calculate
the charge density we integrate the Green function 
along a contour on the complex energy plane which extends
from the bottom of the band up to the Fermi level \cite{vitos00}, using 30 energy points. 
For the Brillouin zone integration we sum up a number of 567 k-points for the 
$Co/Cu$ system and 
839 k points for the $Fe/Cr$ respectively. A cutoff of $l_{max}=8$
for the multipole expansion of the charge density and a cutoff of $l_{max}=4$ for the
wave function was used. The Perdew-Wand \cite{Perdew92} 
parameterization of the local density 
approximation to the exchange correlation potential was used. 
Although the lattice mismatch and relaxation have been shown to influence
the magnetic properties \cite{Wu95} we neglect this effect. 

\subsection{ fcc Co and Co/Cu multilayers} 
\begin{figure} 
\centerline{\psfig{figure=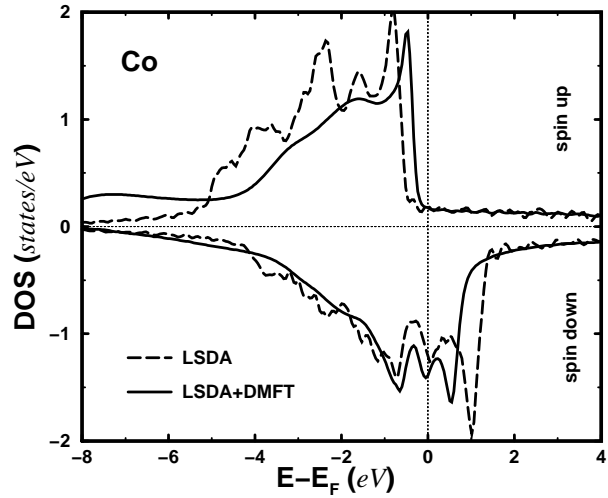,height=3.0in}}
\caption{The LSDA (dashed line) and LSDA+DMFT (solid line) densities of states 
for $fcc$ Co calculated using the EMTO-DMFT method. A significant reduction of 
the exchange splitting can be evidenced.}\label{codos}
\end{figure}

We will first discuss the correlation effects on bulk fcc Co, 
and afterwards investigate the Co/Cu multilayer. The
LSDA and LSDA+DMFT density of states for fcc Co is shown in Fig. \ref{codos}.
The LSDA electronic structure of fcc cobalt is different for the majority and
minority spin electrons, due to it's ferromagnetism $\mu_{LSDA}=1.69 \mu_B$. DOS can 
be characterized qualitatively by a shift of the minority and  majority $d$-bands relative
to each other. Due to this asymmetry the contributions of the two spin channels to the density
of states at the Fermi level will be different, therefore the conductivities are different.
This asymmetry in the DOS is the source of GMR in magnetic Co/Cu multilayers. The most
important feature of the cobalt DOS is that the Fermi level lies above the top of the
$d$-band for the majority spin electrons.

For the LSDA+DMFT calculations a value of the average Coulomb interaction $U=2$ eV 
was chosen. Correlation effects related with spin-flip excitations at non-zero
temperature reduce the magnetic moment, $\mu_{DMFT}=1.42 \mu_B$ in comparison with
the corresponding LSDA value. The asymmetry between the majority/minority
spin channels is kept and an increasing of $12\%$ of DOS at the Fermi level
$N(E_F)$ is evidenced. At low energies the majority spin channel presents a satellite
structure similar to the one found in the Ni. Due to the
perturbation nature of our approach the satellite is shifted towards lower energies. 

\begin{figure}
\centerline{\psfig{figure=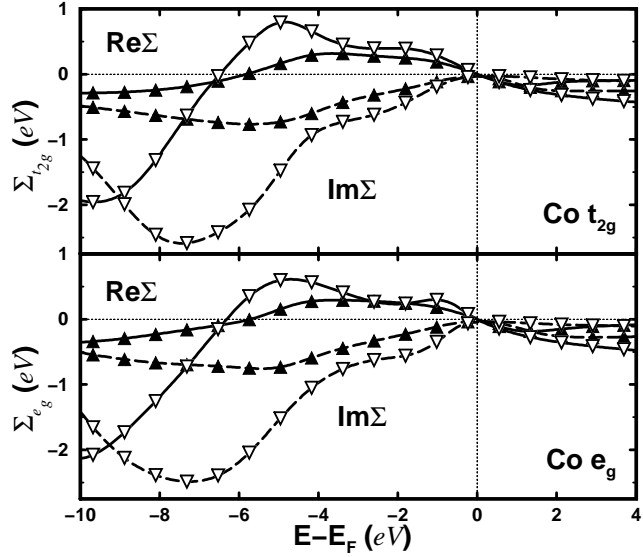,height=3.0in}}
\caption{Spin up (open symbols) and down (closed symbols) self energies for
fcc Co for $t_{2g}$ (upper panel) and $e_g$ (lower panel) orbitals.} \label{cosigm}
\end{figure}

The energy dependence of 
the self-energy for Co is plotted in Fig. \ref{cosigm}. Near the Fermi level 
a typical Fermi liquid behavior is evidenced. For the imaginary part  
$-Im\ \Sigma (E)\propto E^{2}$, meanwhile the real part of the
self-energy has a negative slope $\partial Re\ \Sigma (E)/\partial E<0$, where
$E$ is the electron energy relative to the Fermi level. Due to the fcc structure
Co self-energy shows a considerable similarity to the  
selfenergies of Ni \cite{EMTODMFT}. The  high value of the imaginary part of
self-energy evidenced around $-7$eV in the majority spin channel for both $t_{2g}$ and $e_g$
orbitals produce the satellite visible in DOS Fig. \ref{codos}. 

\begin{figure}
\centerline{\psfig{figure=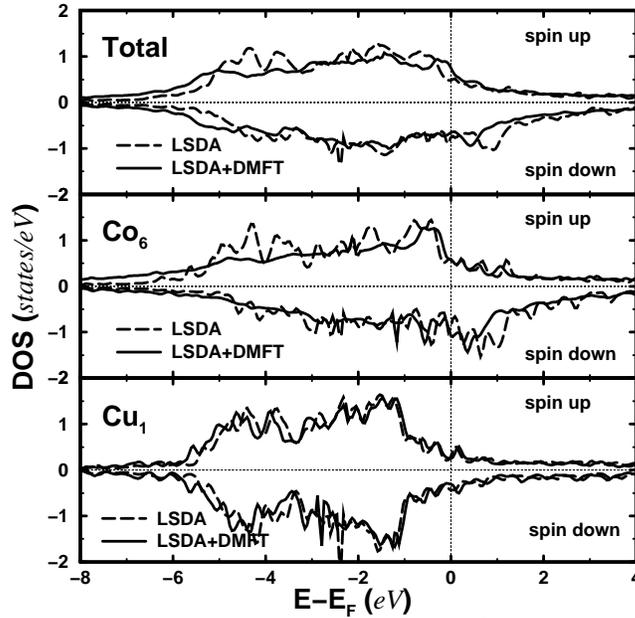,height=3.0in}}
\caption{Total DOS and the interface layers Co$_6/Cu_1$ LSDA (dashed line) and 
LSDA+DMFT (solid line) densities of states.} \label{mldos}
\end{figure}

\begin{figure}
\centerline{\psfig{figure=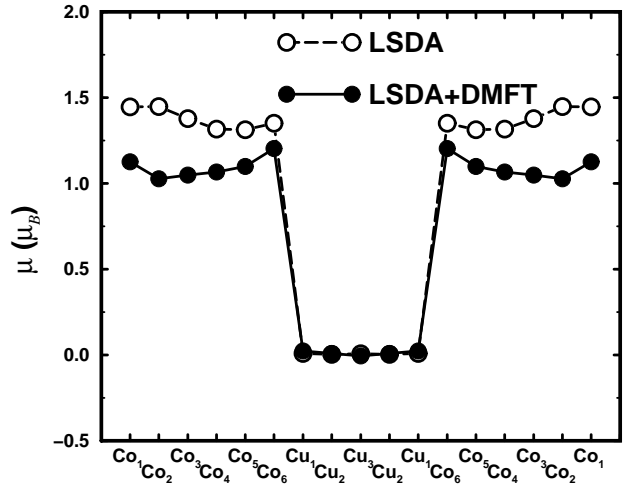,height=3.0in}}
\caption{Distribution of the layer resolved magnetic moment at T=0K (LSDA) and
T=250K (LSDA+DMFT) in $Co/Cu$ superlattice.} \label{mtcocu}
\end{figure}

The $Co_6/Cu_5/Co_5$ multilayer, total and the interface layers DOS, in the ferromagnetic 
orientation are presented in Fig.\ref{mldos}. As we seen in Fig. \ref{mldos}, the DOS 
is asymmetric between the majority and minority spins. In comparison with the 
bulk fcc Co , Fig. \ref{codos}, a qualitative description of DOS of $Co_6/Cu_5/Co_5$ 
can be made, on the base of a "rigid" shift due to the presence of the Cu layers. 
The Co layers magnetic moment increase as we approach the interface, meanwhile the Cu
layers are non-magnetic; the values of the magnetic moments and the 
electronic specific heat coeficients are presented in table \ref{tab1}. 
The latter is given by the relation $\gamma =\pi ^{2}k_{B}^{2}N(E_F)(1+\lambda )/3$
where $N(E_F)$ is the electronic DOS at the Fermi level, $(1+\lambda)$
is the mass enhancement factor caused by the electron-phonon interaction.
The calculated LSDA+DMFT magnetic moments in the $Co_6/Cu_5/Co_5$ superlattice is shown in 
Fig. \ref{mtcocu}, where for comparison we also plot the corresponding quantities
obtained using the LSDA. We note that at finite temperatures the $Cu$ 
layers remain non-magnetic, whereas the value of the magnetic moment on the $Co$ layers 
diminish in comparison with the LSDA one. Even at finite temperatures ($250K$) the 
$Co$ moments couples ferromagnetically across the $Co/Cu$ interfaces.

\begin{figure}
\centerline{\psfig{figure=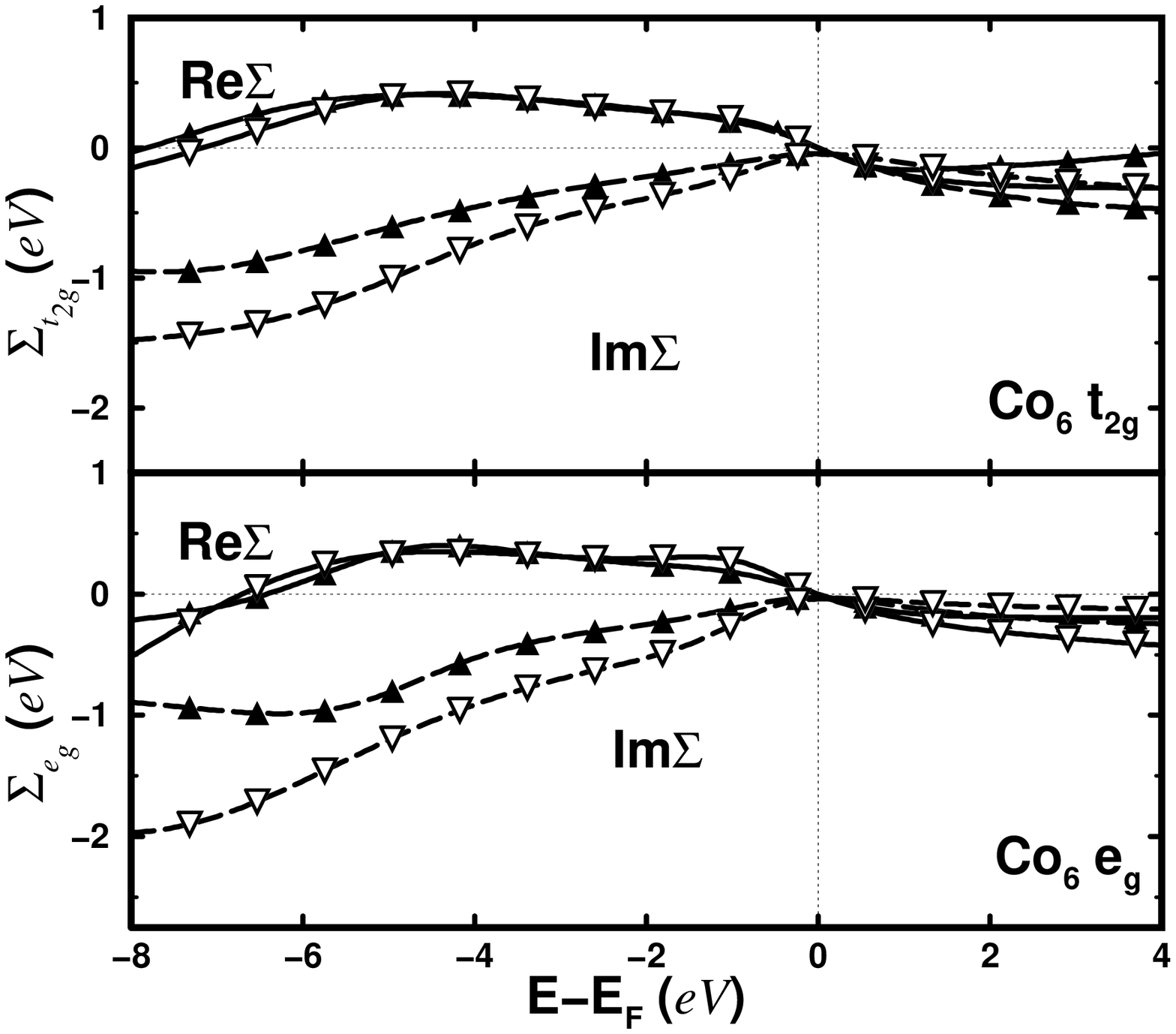,height=3.0in}
\psfig{figure=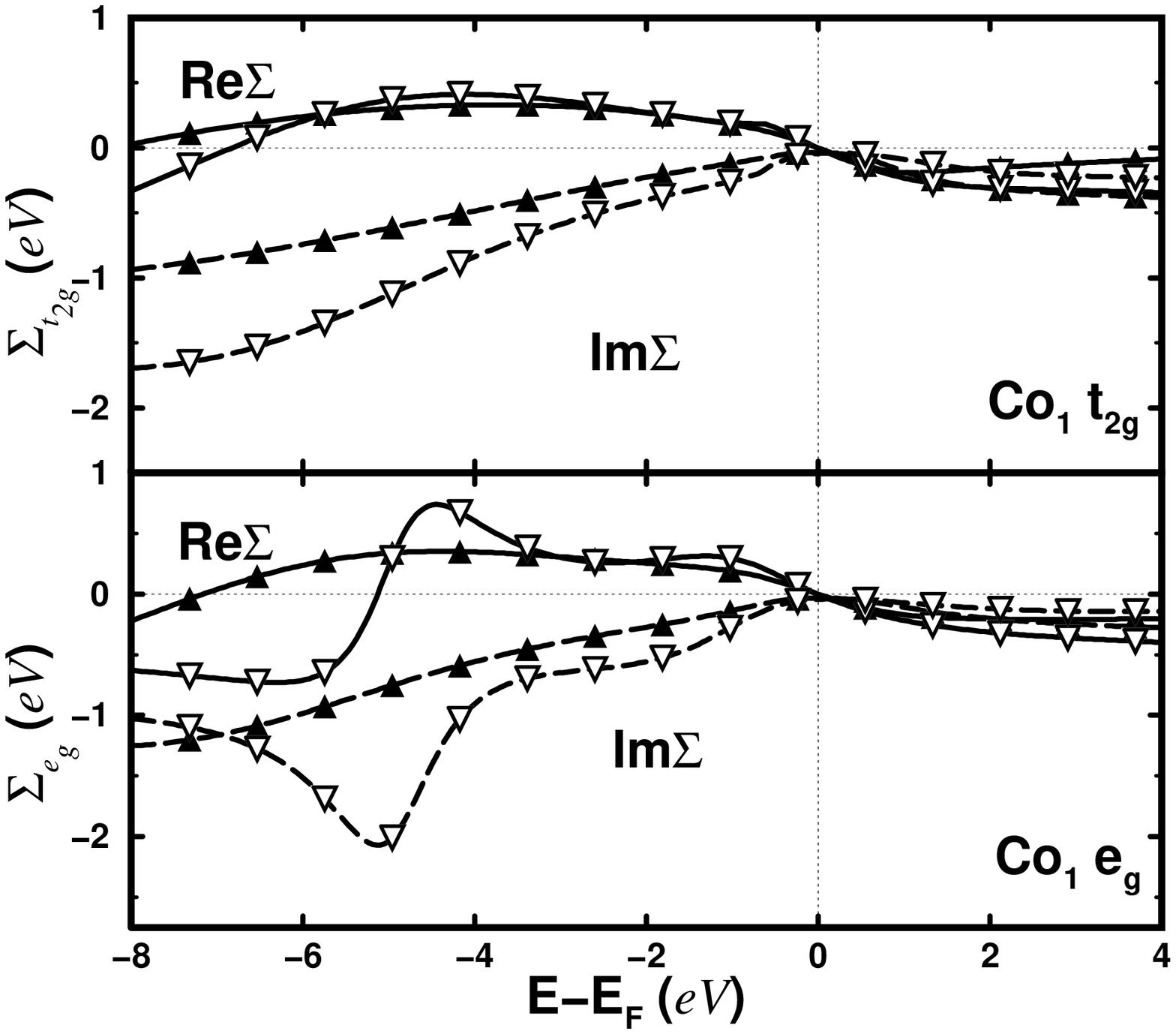,height=3.0in}}
\caption{Layer resolved self-energies: $Co_6$-interface layer and $Co_1$ -
central layer at temperature $T=250$.} \label{co61sigma}
\end{figure}

The energy dependence of the self-energy for the Co interface and central layers 
are plotted in Fig. \ref{co61sigma}.
Near the Fermi level a typical Fermi liquid behavior is evidenced for both the interface
and central Co layers. The imaginary part $-Im\ \Sigma (E)\propto E^{2}$, 
meanwhile the real part 
of the self-energy has a negative slope $\partial Re\ \Sigma (E)/\partial E<0$, where
$E$ is the electron energy relative to the Fermi level. As can be seen the selfenergies
and implicitly the correlation effects are different for the distinct Co layers.

\subsection{ bcc Fe, bcc Cr and Fe/Cr multilayers} 
The correlation effects on bulk bcc Fe, bcc Cr and (001) surface of bcc Fe covered by 
a trilayer of Cr were described in a previous paper \cite{EMTODMFT}. 
The present calculations
addresses the finite temperature and correlations effects in an eight-layer (3Fe+5Cr) bcc
superlattice.

\begin{figure}
\centerline{\psfig{figure=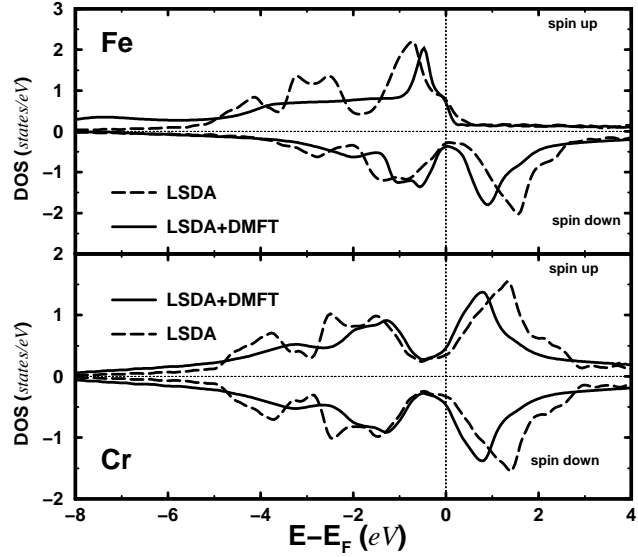,height=3.0in}}
\caption{Bulk bcc iron and chromim DOS, LSDA (dashed line) and
LSDA+DMFT (solid line). The similarities of DOS in the minority spin channel 
determines the minority spin channel behaviour of $Fe_3/Cr_5$ multilayers.} \label{fecrdos}
\end{figure}

In Fig. \ref{fecrdos} the bcc bulk Fe and Cr DOS are presented. As we can see
due to the bcc structure the DOS exhibits a pronounced valley in the middle of
the $d$-bands for both spins. Fermi level lies within the $d$-band for both spin
orientations which provides dominance of the $d$-character. Iron is magnetic 
$\mu_{LDA}=2.25 \mu_B$, whereas chromium is non-magnetic.

\begin{figure}
\centerline{\psfig{figure=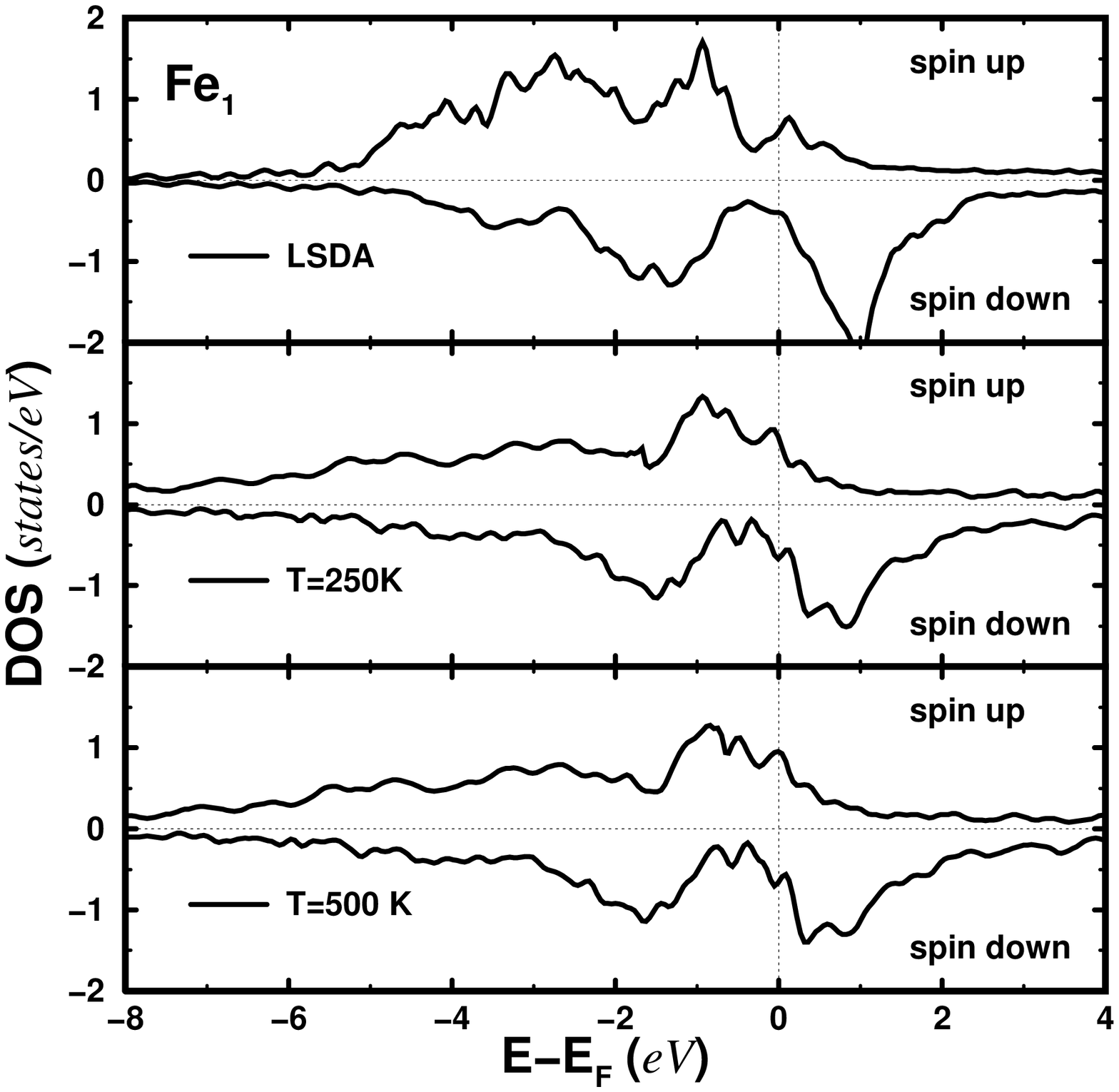,height=3.0in}
\psfig{figure=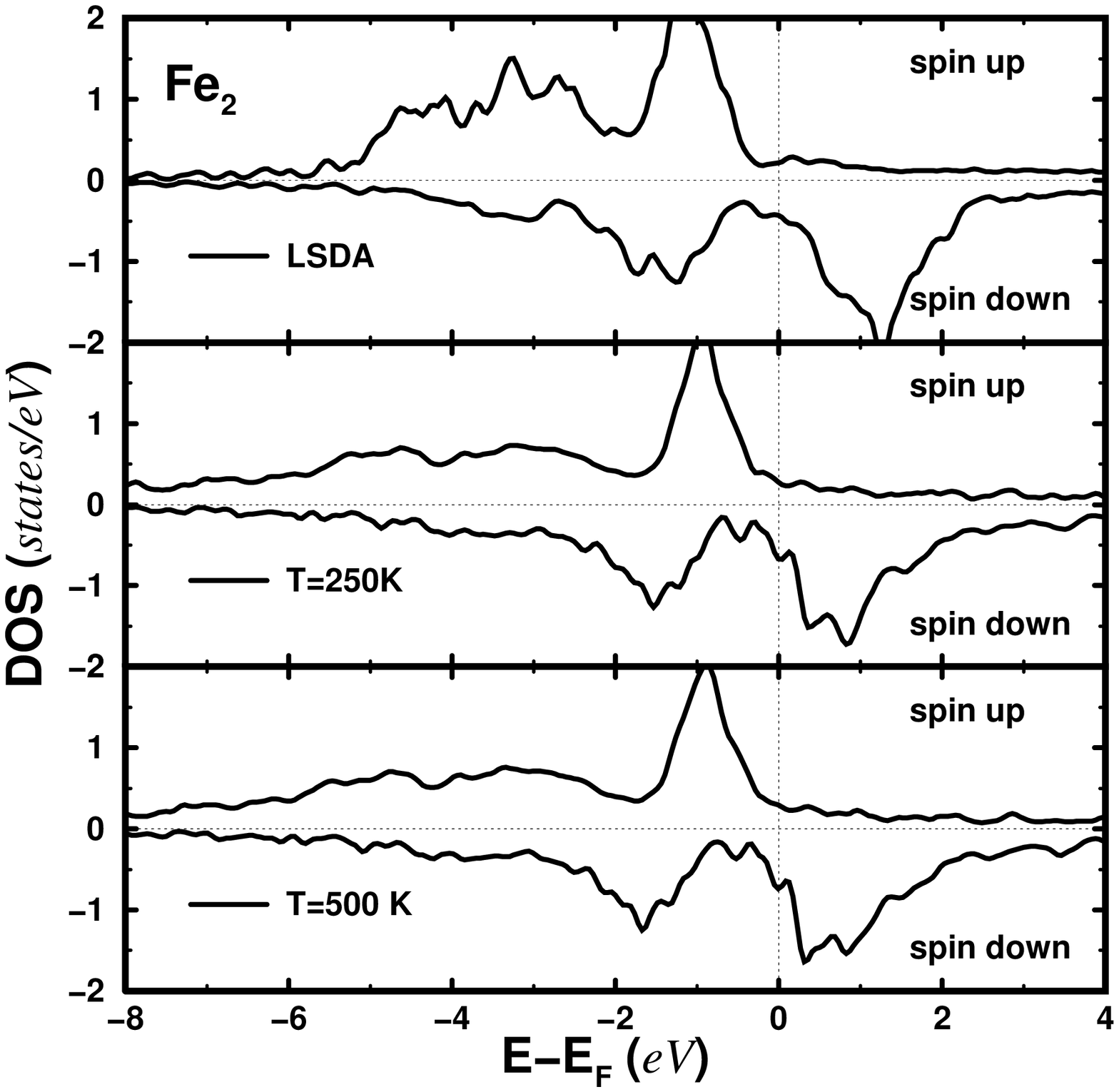,height=3.0in}}
\caption{Total DOS for both types of Fe layers: $Fe_1$-interface layer and $Fe_2$ -   
central layer. The temperature dependence of DOS is presented for $T=0, 250$ 
and $500K$ respectively.} \label{fe12mldos} 
\end{figure}

Similarly to the bulk DOS, 
as it is evident from Fig. \ref{fe12mldos} the Fermi level in the Fe/Cr system
lies within the $d$-band for both spin orientations. DOS exhibits a
pronounced valley for the minority spins with the Fermi level lying almost at
the bottom of this valley. This feature of the band structure is a consequence of the 
similarity of the DOS of minority spin electrons of iron and the minority 
spin channel DOS of chromium Fig. \ref{fecrdos}.

\begin{figure}
\centerline{\psfig{figure=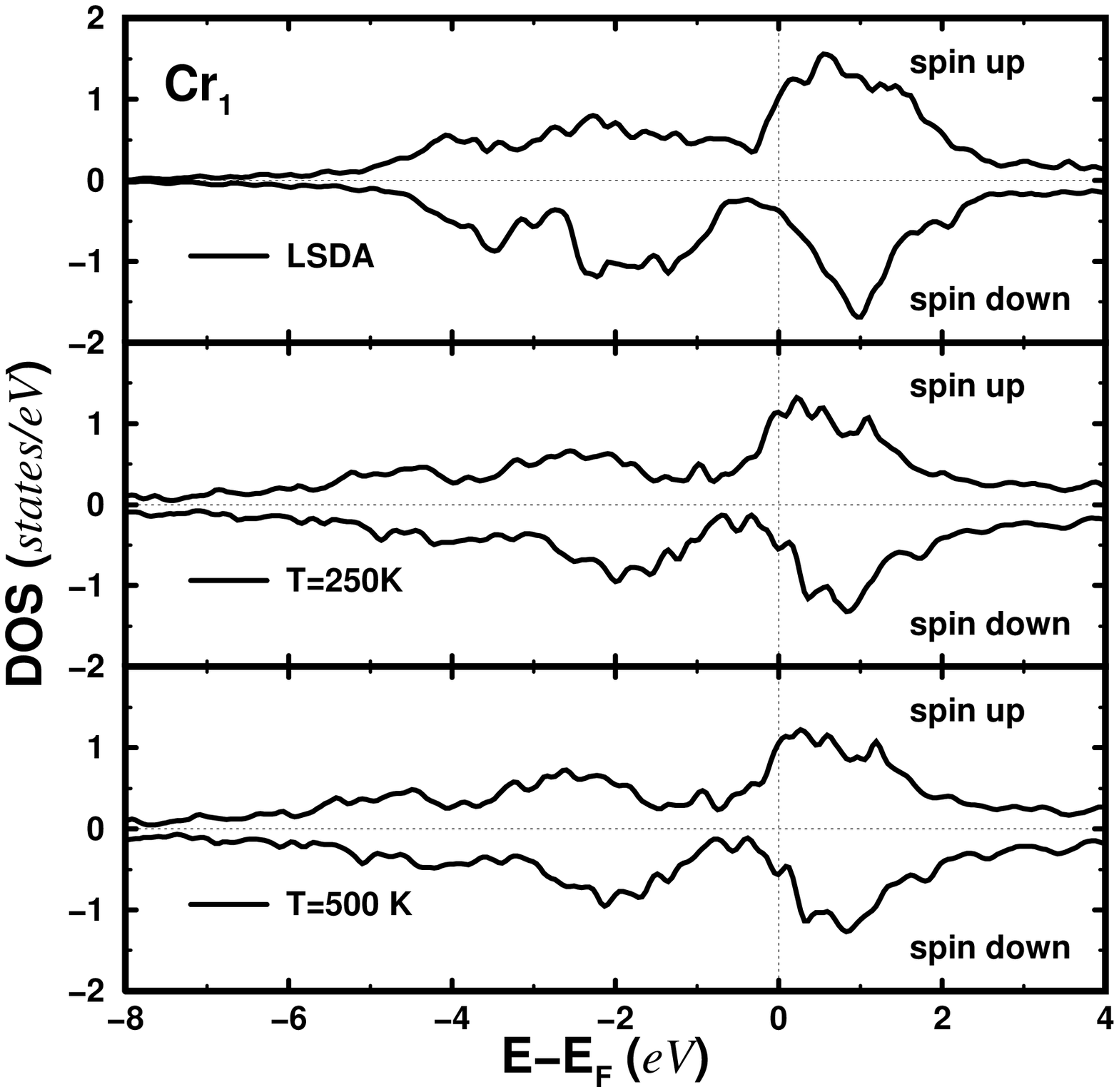,width=2.0in}
\psfig{figure=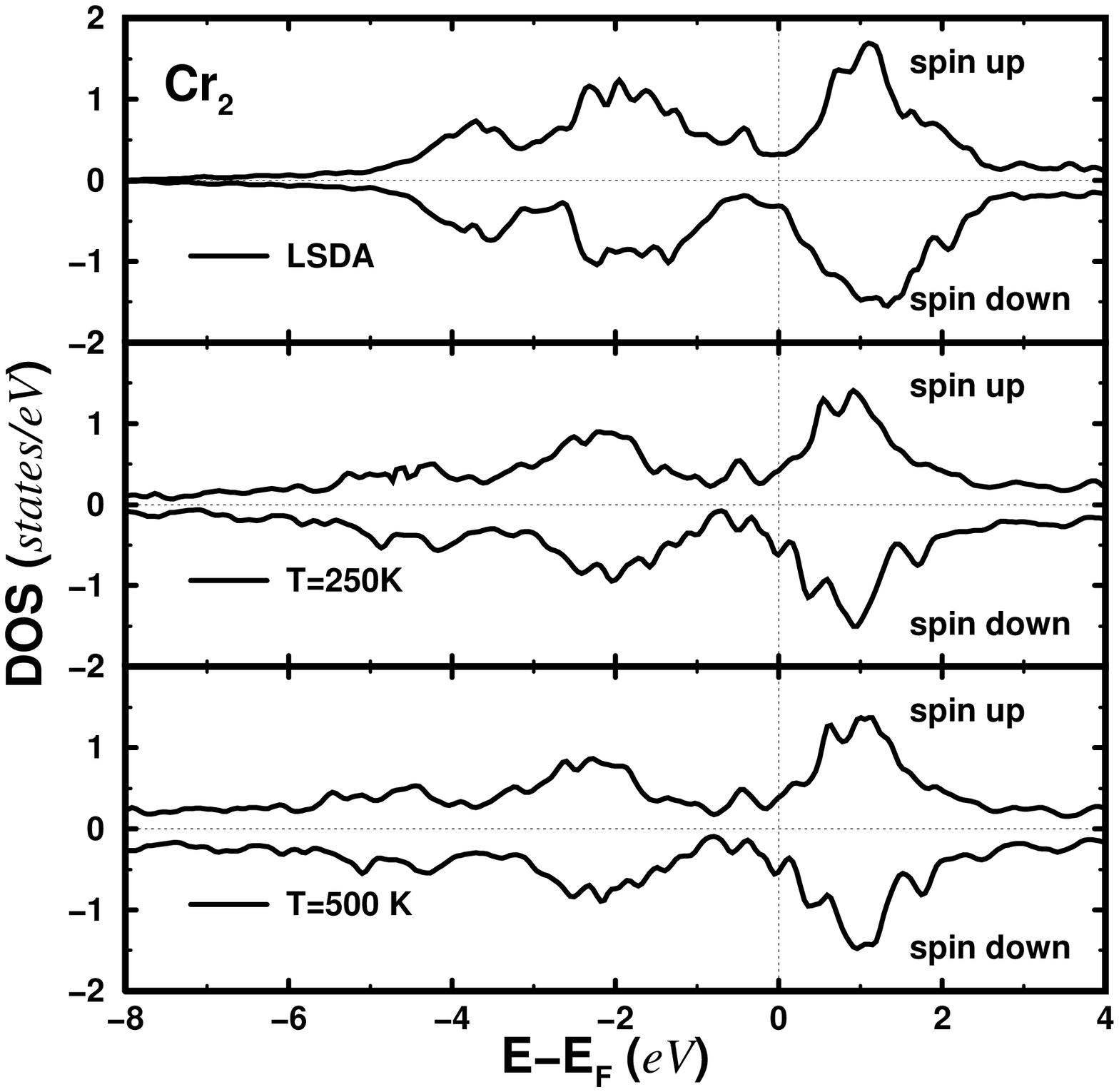,width=2.0in} 
\psfig{figure=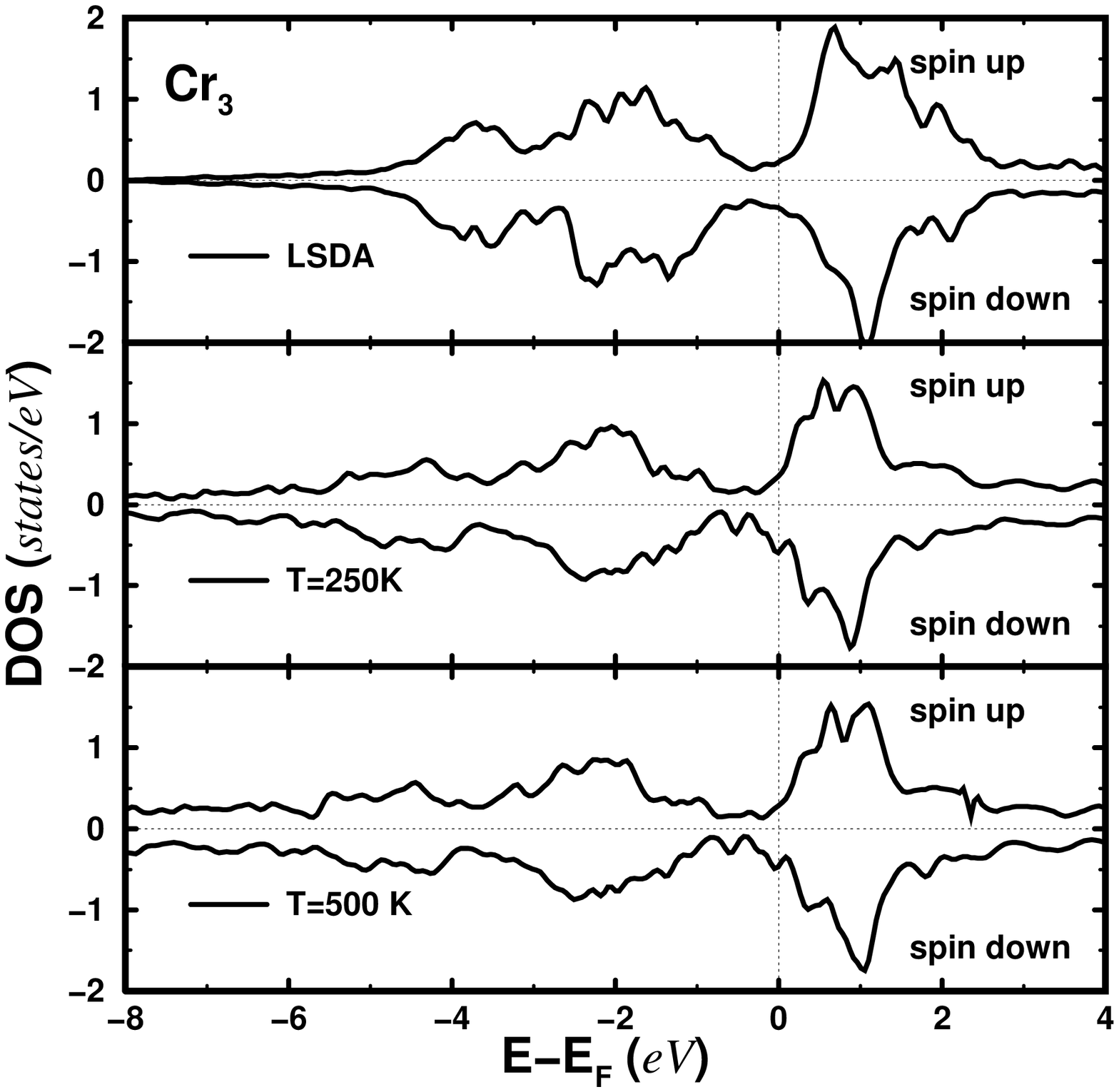,width=2.0in}}
\caption{Total DOS for the three types of Cr layers: $Cr_1, Cr_2$ - interface layers and $Cr_3$ -   
central layer. The temperature dependence of DOS is presented for $T=0, 250$ 
and $500K$ respectively.} \label{cr123mldos}
\end{figure}

Many-body effects described in the framework of the LDA+DMFT, were investigated 
for different temperatures: $T=250, 500K$. Since we expect similar correlation 
effects in the Fe/Cr system the same value of the average Coulomb interaction
$U=2eV$ and the same exchange correlation energy $J=0.9eV$ was chosen. The DOS
for the Fe layers and Cr layers are presented in Fig. \ref{fe12mldos} and 
Fig. \ref{cr123mldos} respectively. 

In the case of Fe, Fig. \ref{fe12mldos} the correlation effects are manifested in a 
slightly different way 
for central and interface Fe layers. One can see that for the interface ($Fe_1$) layer  
the LSDA peack of the unoccupied DOS close to the Fermi level, is pinned at the Fermi level 
producing an enhancement of $N(E_F)$. The spectral weight of the main peak in the occupied 
part ($-1eV$) of spin up channel, is transfered closer to the Fermi level, as temperature 
is increased. In the same time finite temperature effects smear out the low energy 
features of DOS, situated in the energy range of $-4, -2$ eV. In the spin down channel, 
as the correlation effects are switched on a peak at the Fermi level appears. The spectral weight
of the $1eV$ spin down LSDA peak is transfered towards the Fermi level and a new peak
appears around $0.25eV$. At $T=250K$, Fig. \ref{fe12mldos}, the weight of the
former $1eV$ peak dominates the weight of the $0.25eV$ peak, but as the
temperature increases, $T=500K$, more spectral weight is transfered to the $0.25eV$ peak.

Similarly to the interface $Fe_1$ layer, the spin down density of 
states of the central $Fe_2$ layer 
show the appearance of the peak at the Fermi level, and the temperature dependence of the 
spectral weight transfer towards the Fermi level. Due to the correlation effects, 
the narrowing of the width of the $-1eV$ peak in the spin up channel is evidenced. As the 
temperature is increased, the $-1eV$ peak shows a slight shift towards 
the Fermi energy, but it's 
width is not significantly changed. 

\begin{figure}
\centerline{\psfig{figure=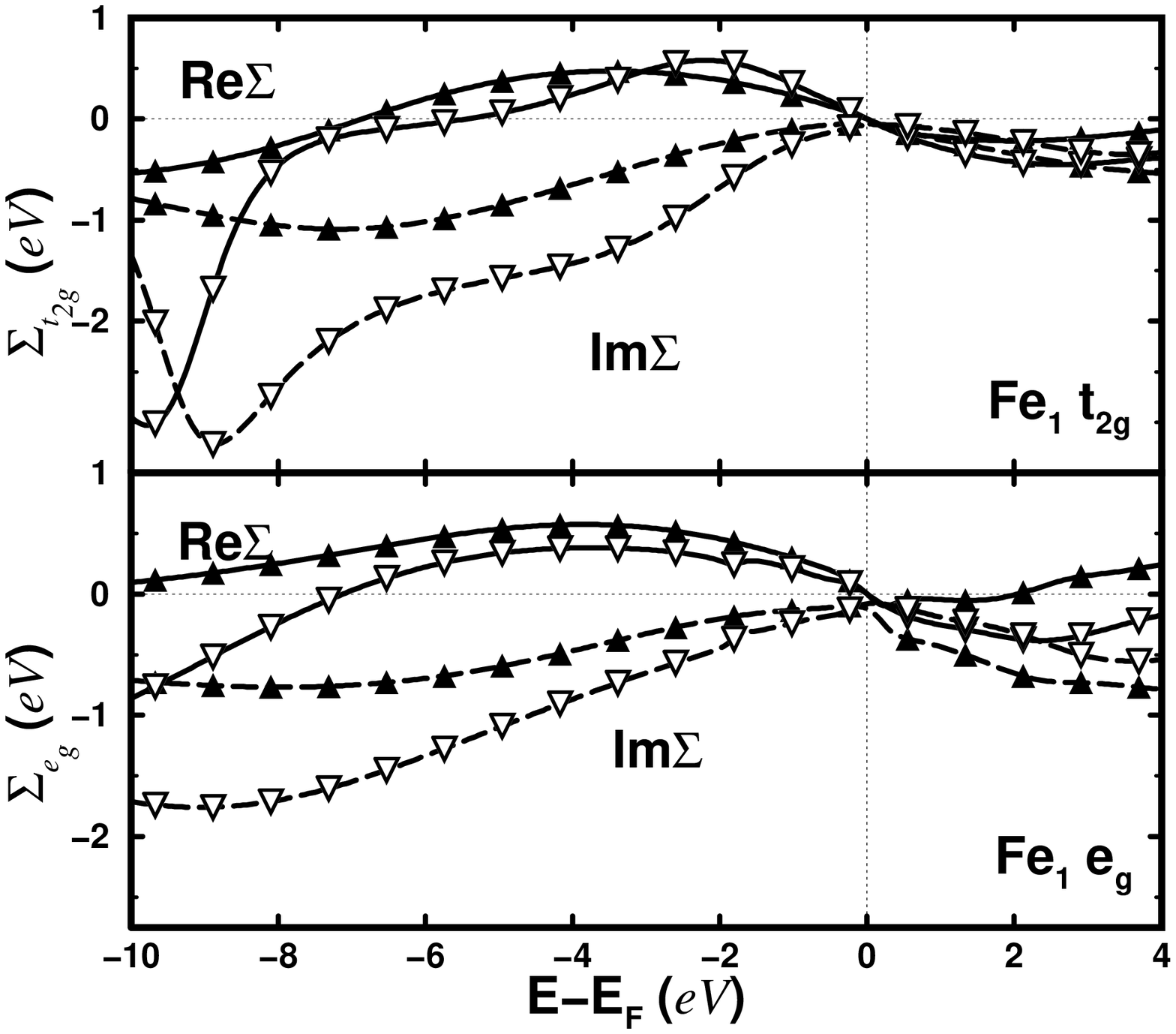,height=3.0in}
\psfig{figure=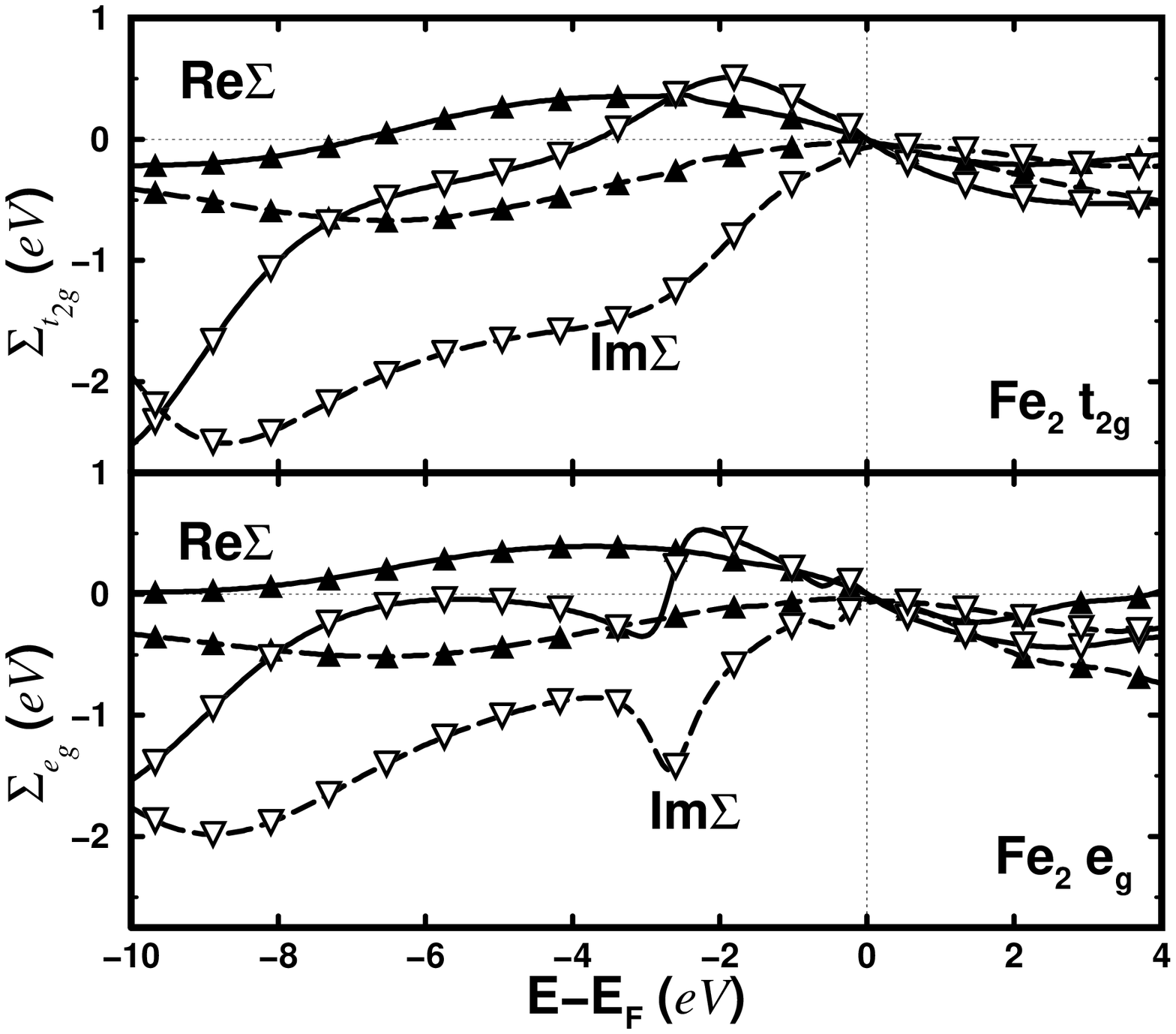,height=3.0in}}
\caption{Layer resolved self-energies: $Fe_1$-interface layer and $Fe_2$-
central layer at temperature $T=500$.} \label{fe12sigma}
\end{figure}

The energy dependence of the self-energy for the Fe layers 
are plotted in Fig. \ref{fe12sigma}.
Near the Fermi level a typical Fermi liquid behavior is evidenced for both the interface
and central Fe layers. The imaginary part $-Im\ \Sigma (E)\propto E^{2}$, 
meanwhile the real part 
of the self-energy has a negative slope $\partial Re\ \Sigma (E)/\partial E<0$, where 
$E$ is the electron energy relative to the Fermi level. As can be seen the selfenergies 
and implicitly the correlation effects are different for the two distinct Fe layers. 

All the above correlation effects can be recognized in the 
Cr layers as well Fig. \ref{cr123mldos},
in particular the formation of the peak at the Fermi level being a significant feature
for the spin down channel. 
According to the LSDA calculation the density of states of the central Cr
layer ($Cr_3$) is almost the same as that of the bcc Cr bulk Fig.\ref{fecrdos}.
As going from the interface Cr layers ($Cr_1$ and $Cr_2$) towards the central layer
($Cr_3$) the spin up channel density of states show the formation of a valley in the
DOS near the Fermi level characteristic to the bulk behavior Fig. \ref{cr123mldos}.
The interface, $Cr_1$ layer, DOS is modified appreciable because of the presence of
nearby strongly ferromagnetic $Fe$ layer. It is important to note, however that
the spin down channel of all $Cr$ layers are not significantly affected by the
proximity to the magnetic $Fe$ layer.
The spectral weight transfer is also present, but for the interface
$Cr_1$ layer is not that evident since hybridization with the nearby 
strongly ferromagnetic $Fe$ layer is present.

\begin{figure}
\centerline{\psfig{figure=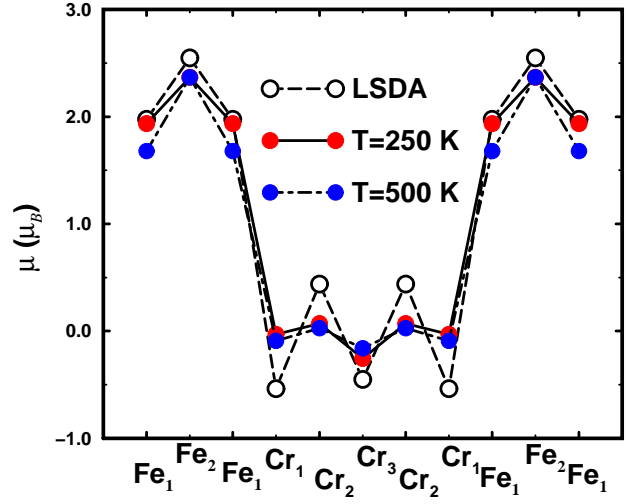,height=3.0in}}
\caption{The distribution of the local magnetic moments at T=0, 250 and 500K 
in 3Fe +5 Cr superlattice as given by the LSDA and the different DMFT calculations}
 \label{fecrmom}
\end{figure}

\begin{figure}
\centerline{\psfig{figure=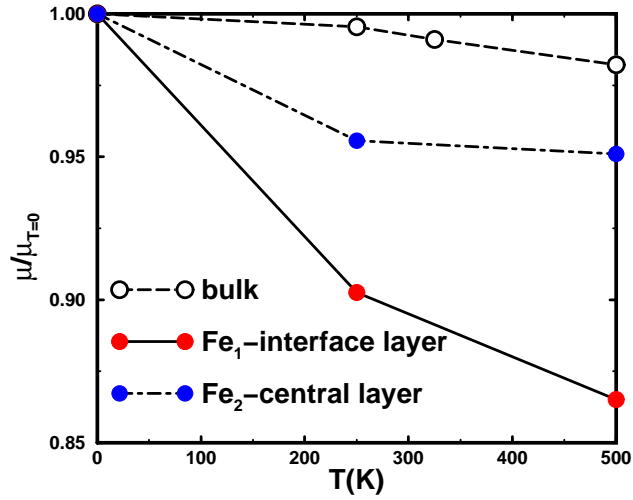,height=3.0in}}
\caption{ Temperature dependence of the calculated magnetic moments 
on the Fe layers is presented in comparison the temperature dependence of the
calcualted bulk bcc Fe magnetic moments. The value at 
$T=325K$ is taken from the reference [19].
The normalization of the magnetic moment is done with respect to the LSDA value
($\mu_{T=0}$).} \label{mtfe}
\end{figure}

\begin{figure}
\centerline{\psfig{figure=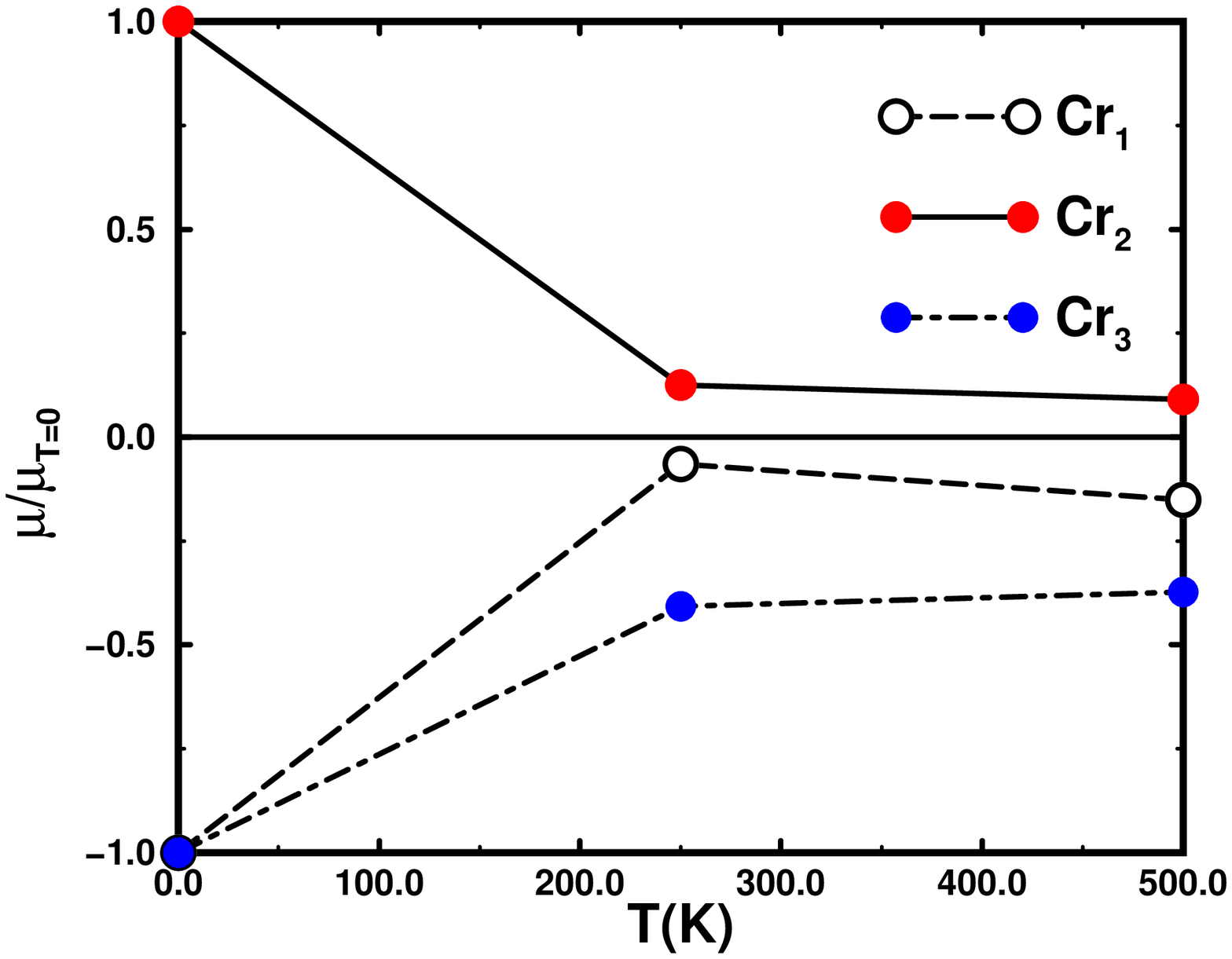,height=3.0in}}
\caption{Temperature dependence of the magnetic moment on the Cr layers. The
normalization of the magnetic moment is done with respect to the LSDA value 
($\mu_{T=0}$). }\label{mtcr}
\end{figure}

In Fig. \ref{fecrmom} we notice that the magnetic moments of Cr layers
alternate from layer to layer and the Fe moments couple ferromagnetically 
across the Fe/Cr interfaces. According to our LDA+DMFT calculation this Fe/Cr antiparallel 
coupling across the interface was estimated to be more stable compared to the 
parallel solution. The temperature dependence of the $Fe$ magnetic moment on each 
layer  is presented in Fig. \ref{mtfe}. The central $Fe_2$ layer 
follows approximately the behavior of the bulk values, meanwhile the 
interface layer $Fe_1$ has a faster temperature decrease. On the other hand the $Cr$ 
layers show a very peculiar temperature dependence of the magnetic moment due to the couplings
across the Fe/Cr layers. Fig. \ref{mtcr} display the Cr layers temperature 
dependence of the magnetic
moment which shows a deviation from the Brillouin function.  

\begin{figure}
\centerline{\psfig{figure=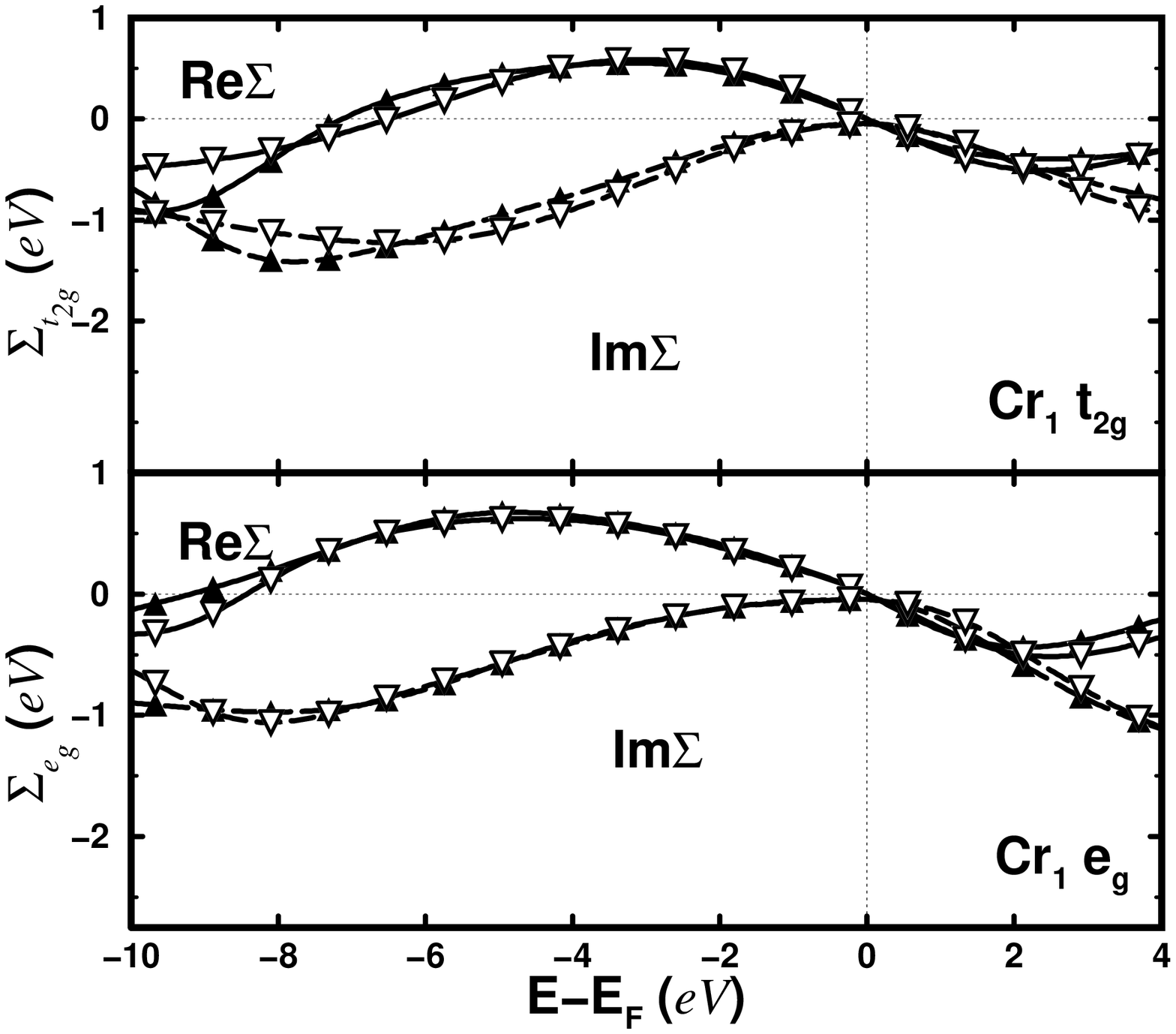,width=2.0in}
\psfig{figure=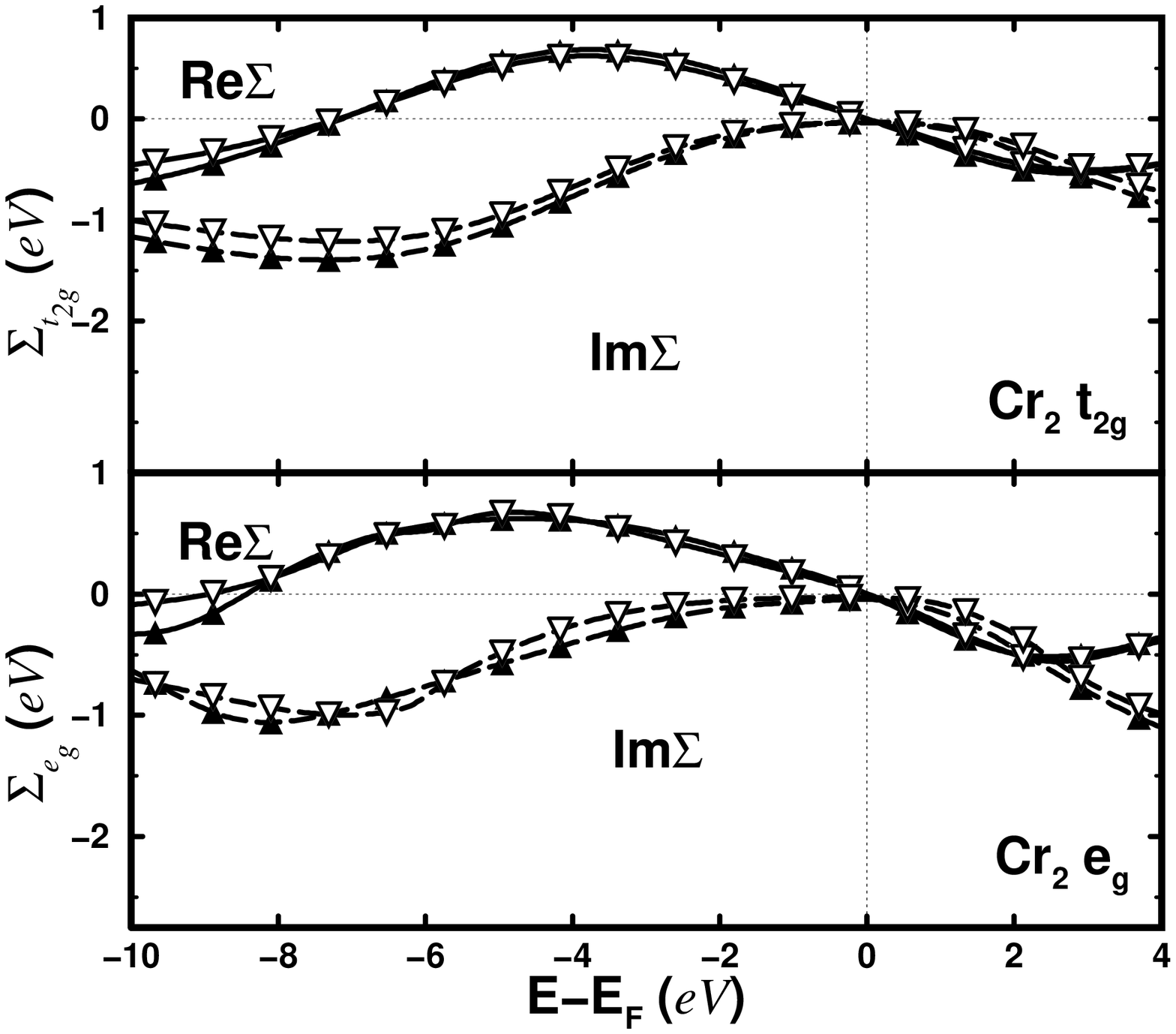,width=2.0in}
\psfig{figure=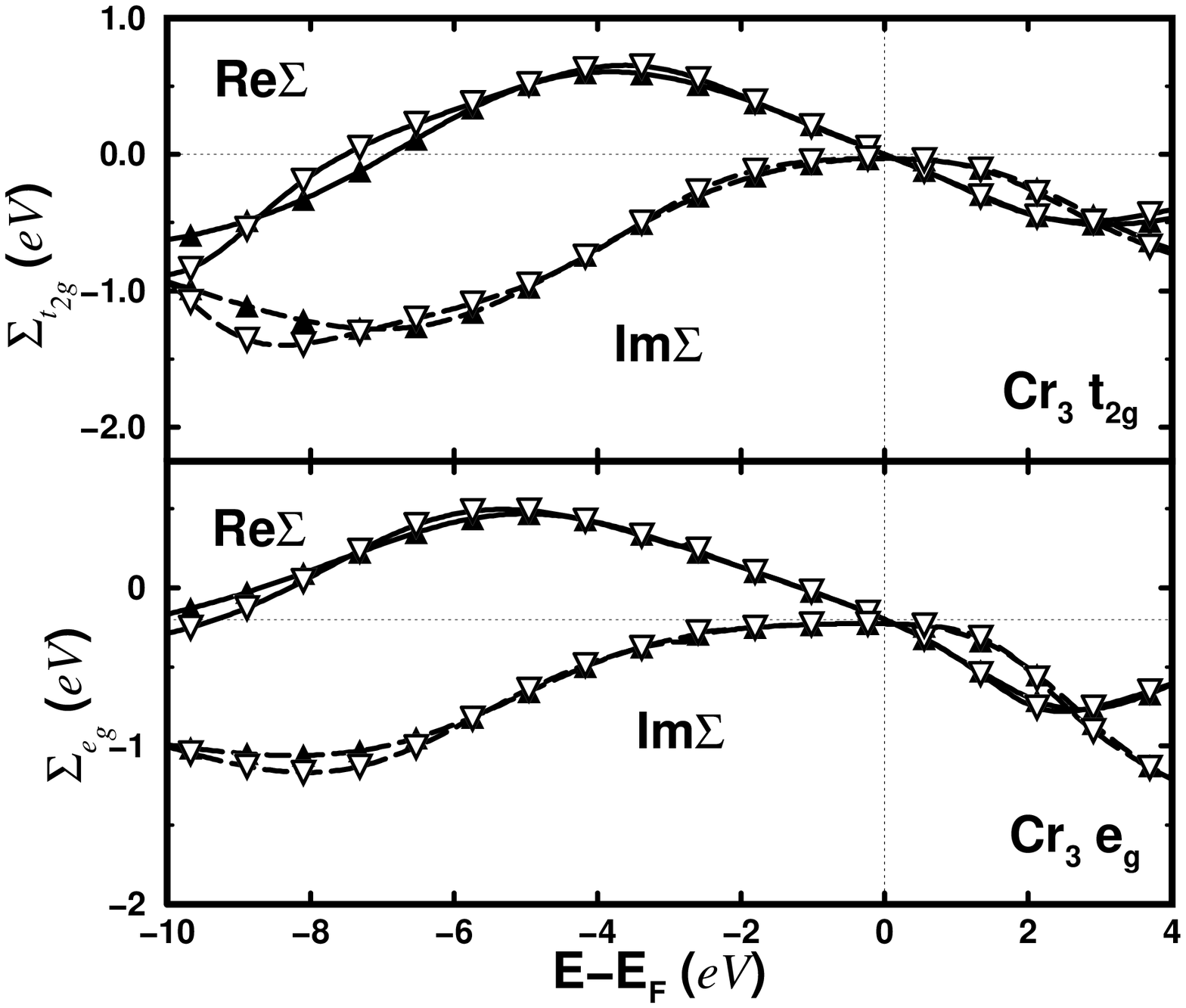,width=2.0in}}
\caption{Layer resolved self-energies for Cr layers: $Cr_1, Cr_2$ - interface layers and $Cr_3$ -
central layer for for $T=500$ .} \label{cr123sigma}
\end{figure}

The $Cr$ layers being non-magnetic the majority and minority spin self-energies 
Fig.\ref{cr123sigma} are identical. However correlation effects seems to be more important for 
the central $Cr_3$ layer. 

\section{Conclusions} 
\label{conclusion} 

The mechanism of giant magnetoresistence in magnetic multilayers 
is usually related to the spin-dependence of the scattering
process. The spin dependent scattering is assumed to arise from 
spin-dependent random potentials produced by magnetic impurities at
the interface or in the bulk of the ferromagnetic layers 
\cite{Levy90,Itoh95}. Recently an improved prediction of the 
GMR was obtained by combining the disorder effects with the accurate
spin-dependent electronic structure calculations \cite{Tsymbal96}. 
However finite temperature properties and correlations effects were not 
taken into account.

Several theoretical approaches have been used to explain the magnetic
properties of such superlattice structures. Many of these approaches are
based on the RKKY-like model \cite{Yafet,Wang}, tight-binding models \cite
{Stoeffler} and, recently, on the results of  {\it ab initio} electronic
structure calculations \cite{Ounadjela}. The magnetic coupling studied in the
framework of these models was shown to result from the interplay between the
direct $d-d$ hybridization of Fe and Cr atoms and indirect exchange through
the $sp$ electrons. The $sp-d$ coupling \cite{Ounadjela} was found to be
reminiscent of the RKKY interaction only for superlattices with more than
four $Cr$ layers \cite{Ounadjela}. However, for the case of three $Cr$ layers the $sp-d$
coupling model cannot explain the ferromagnetic ordering of $Fe$ atoms.

Based on the theory of spin-fluctuations developed by Moriya \cite{Moriya85} 
and applied by Hubbard \cite{Hubbard79} for the case of several transition metals,
Hasegawa \cite{Hasegawa88} showed that spin-fluctuations  plays an important
role in discussing the temperature dependence of the GMR. Due to the static 
approximation employed in the model \cite{Hasegawa88}, dynamical effects of the spin-fluctuations
were neglected, therefore any definite conclusion on the temperature dependence
of GMR were not drawn. Although the explicit calculation  of the resistivity and GMR
is not the purpose of the present paper, our results include dynamical correlations
described in the framework of DMFT \cite{Georges96}, being a promising starting point for
a direct evaluation of the GMR in multilayer systems. 

Using an first principle LDA+DMFT approach \cite{EMTODMFT} we examined
the correlation effects and the finite temperature magnetic properties of
some $Co/Cu$ and $Fe/Cr$ superlatticels. Our calculations evidenced
a peculiar temperature dependence of the magnetic moments near the
interface. The correlation effects proved to be different for different atomic layers.
Based on the spin polarized T-matrix FLEX, DMFT solver,
the correlation effects capture the spin fluctuations that plays 
primary roles at finite temperatures. Recent experiments on Fe/Cr trilayers
showed the existence of magnetic fluctuations \cite{Kentzinger03}. Even though
the magnetic excitation were attributed to structural
and magnetic disorder in the vicinity of both Fe/Cr interface \cite{Kentzinger03}, 
results presented in this paper suggest that electron-electron interactions
give rise to magnetic excitations, which are common features in 3d transition
metals systems. 

It is worthwhile to emphasize that the enhanced DOS at the Fermi level,
having a many-body correlation origin, can play an important role in the
GMR, since this DOS enhancement is strongly spin-dependent. It is more effective 
in the minority channels of $Co/Cu$ And $Fe/Cr$ systems,
giving the result of a quasiparticle peak
centered at the Fermi level. Our calculation is in good agreement with the
tendency of the enhancement of electronic contribution in $Fe/Cr$ magnetic
multilayers \cite{Revaz02}.

\section*{Acknowledgments} 
 
L.C. acknowledges the financial supports from: Uppsala University,  the 
''Computational Magnetoelectronics'' RTN project (HPRN-CT-2000-00143).  
This work was supported by the Netherlands 
Organization for Scientific Research (NWO), grant NWO 047-008-16. Discussions with
M.I. Katsnelson, O. Erikksson and I.A. Abrikosov are acknowledged.

\begin{table}[tbp] 
\caption{Theoretical magnetic moments $\mu$ and electronic specific heat coeficients $\gamma$ 
calculated for bulk bcc Fe, Cr and fcc Co. The experimental values for $\gamma$ include the 
contribution from the electron-phonon coupling, which is not included in the results obtained from
band structure calculations. In the last three columns, the parameters used in the self-consistent 
LSDA+DMFT calculations are listed.}
\label{tab1}  
\begin{tabular}{ccccccccc} 
& $\mu_{LDA}$ & $\mu_{DMFT}$ & $\gamma_{LDA}$ & $\gamma_{DMFT}$ & $% 
\gamma_{exp}$ & $T$ & $U$ & $J$ \\  
& $(\mu_B)$ & $(\mu_B)$ & $(mJ/K^2\ mol)$ & $(mJ/K^2 \ mol)$ & $(mJ/K^2\ 
mol) $ & $(K)$ & $(eV)$ & $(eV)$ \\ \hline\hline 
Co & 1.62 & 1.38 & 5.43 & 6.78 &  -  & 250 & 2 & 0.9 \\  
Fe & 2.25 & 2.24 & 2.43 & 2.61 & 3.11,3.69$^a$& 250 & 2 & 0.9 \\  
Cr & - & - & 1.73 & 2.40 & 3.5$^b$ & 250 & 2 & 0.9 
\end{tabular} 
$^a$ Ref.\ \cite{phillips71}             \\
$^b$ Non-magnetic Cr, Ref.\ \cite{fawcett88}. \\
\end{table}

\begin{table}[tbp]
\caption{The calculated magnetic moments in the LSDA (T=0K) and LSDA+DMFT (T=250K)
for the $Co_6/Cu_5/Co_5$ superlatice. }
\label{tab2}
\begin{tabular}{cccccccc}
& $\mu_{LDA}$ & $\mu_{DMFT}(T=250K)$ & $U$ & $J$ \\
& $(\mu_B)$ & $(\mu_B)$  & $(eV)$ & $(eV)$ \\ \hline\hline
$Co_1(Co/Cu)$ & 1.44 & 1.12 & 2.0 & 0.9 \\  
$Co_2(Co/Cu)$ & 1.44 & 1.02 & 2.0 & 0.9 \\  
$Co_3(Co/Cu)$ & 1.37 & 1.04 & 2.0 & 0.9\\ 
$Co_4(Co/Cu)$ & 1.31 & 1.06 & 2.0 & 0.9\\ 
$Co_5(Co/Cu)$ & 1.31 & 1.09 & 2.0 & 0.9\\ 
$Co_6(Co/Cu)$ & 1.35 & 1.20 & 2.0 & 0.9\\ 
$Cu_1(Co/Cu)$ & 0.00 & 0.02 & 0.0 & 0.0\\ 
$Cu_2(Co/Cu)$ & 0.00 & 0.00 & 0.0 & 0.0\\ 
$Cu_3(Co/Cu)$ & 0.00 & 0.00 & 0.0 & 0.0
\end{tabular} 
\end{table} 

\begin{table}[tbp]
\caption{Temperature dependence of layer resolved magnetic moments. The bulk
values of the magnetic moment for the same temperatures, and the parameters
used in DMFT calculations are presented in the table.}
\label{tab3}
\begin{tabular}{cccccc}
& $\mu_{LDA}$ & $\mu(T=250K)$ & $\mu(T=500K)$ &  $U$ & $J$ \\
& $(\mu_B)$   & $(\mu_B)$    & $(\mu_B)$    & $(eV)$ & $(eV)$ \\ \hline\hline
$Fe  (bulk)$ & 2.25  & 2.24 & 2.21 &  2.0 & 0.9 \\
$Fe_1(Fe/Cr)$ & 1.97  & 1.78 & 1.70 &  2.0 & 0.9 \\
$Fe_2(Fe/Cr)$ & 2.54  & 2.38 & 2.42 &  2.0 & 0.9 \\
$Cr_1(Fe/Cr)$ &-0.54  &-0.04 &-0.08 &  2.0 & 0.9\\
$Cr_2(Fe/Cr)$ & 0.43  & 0.05 & 0.04 &  2.0 & 0.9\\
$Cr_3(Fe/Cr)$ &-0.45  &-0.18 &-0.16 &  2.0 & 0.9
\end{tabular}
\end{table}

\begin{references} 

\bibitem{GMR}  A. Fert, Phys. Rev. Lett. {\bf 61}, 2472 (1988); P.Gr\"{u}% 
nberg, Phys. Rev. B {\bf 39}, 4828 (1989). 

\bibitem{Schep95} K.M. Schep, P. J. Kelly and G.E.W. Bauer, \prl {\bf 74}
, 568 (1995). 

\bibitem{Butler93} W.H. Buttler, J.M. MacLaren and X.G. Zhang, Mater. Res. Soc. Symp. 
Proc. {\bf 313}, 59 (1993). 

\bibitem{Nesbet94} R.K. Nesbet,  J. Phys.
Condens. Matter {\bf 6}, L449, (1994). 
 
\bibitem{Zahn95} P. Zahn, I. Mertig, M. Richter and H. Eschrig \prl {\bf 75}, 
2996 (1995).

\bibitem{Tsymbal97} E.Yu. Tsymbal and D.G. Pettifor, J. Appl. Phys  {\bf 81}, 
4579 (1997); Solid State Phys {\bf 56}, 113 (2001).

\bibitem{Weinberger2003} P. Weinberger, Physics Reports-Review Section of Physics Letters, 
{\bf 377}, 281 (2003).

\bibitem{Katsnelson99}  M. I. Katsnelson and A. I. Lichtenstein J. Phys.
Condens. Matter {\bf 11}, 1037, (1999).

\bibitem{Co}  S. Monastra et al, Phys. Rev. Lett. {\bf 88}, 236402 (2002).

\bibitem{Lichtenstein01}  A. I. Lichtenstein, M. I. Katsnelson, and G.
Kotliar, Phys. Rev. Lett. {\bf 87}, 067205 (2001).

\bibitem{Antropov}M. van Schilfgaarde, V.P. Antropov,
J. Appl. Phys. {\bf 85}, 4827 (1999).

\bibitem{Schad95} R.Schad, C.D. Potter, P. Belien, G. Verbach, J. Dekoster, G. Langouche,
V.V. Moschalkov and Y. Bruynseraede, J. Magn. \& Magn. Mater., {\bf 148}, 331 (1995);
R.Schad, C.D. Potter, P. Belien, G. Verbach, V.V. Moschalkov and Y. Bruynseraede,
Appl. Phys. Lett., {\bf 64}, 3500 (1994). 

\bibitem{Parkin91} S.S.P. Parkin, Z.G. Li and D.J. Smith, Appl. Phys. Lett., {\bf 58},
2710 (1991).

\bibitem{Tsymbal96} E.Yu. Tsymabl and D.G. Pettifor,  J. Phys.
Condens. Matter {\bf 8}, L569, (1996).

\bibitem{Revaz02}  B. Revaz el al., Phys. Rev. B {\bf 65}, 094417 (2002).

\bibitem{Kulikov97}  N. I. Kulikov and C. Demangeat, Phys. Rev. B {\bf 55},
3533 (1997).

\bibitem{Georges96}  A. Georges, G. Kotliar, W. Krauth, and M. J. Rozenberg,
Rev. Mod. Phys. {\bf 68}, 13 (1996).

\bibitem{DMFT} V. I. Anisimov, A. I. Poteryaev, M. A. Korotin, A. O. Anokhin,
G. Kotliar, J. Phys.: Condens. Matter {\bf 9}, 7359 (1997); 
A. I. Lichtenstein and M. I. Katsnelson,
\prb {\bf 57}, 6884 (1998). 

\bibitem{EMTODMFT} L. Chioncel, L. Vitos, I. A. Abrikosov, 
J. Koll\'ar, M. I. Katsnelson, and A. I.
Lichtenstein, Phys. Rev. B. {\bf 67}, 235106 (2003).

\bibitem{andersen94}  O. K. Andersen, O. Jepsen, and G. Krier, in {\it %
Lectures on Methods of Electronic Structure calculations}, edited by V.
Kumar, O.K. Andersen, and A.Mookerjee (World Scientific Publishing Co.,
Singapore, 1994), p. 63; O. K. Andersen and T. Saha-Dasgupta, \prb {\bf 62},
R16219 (2000).

\bibitem{vitos00}  L. Vitos, H. L. Skriver, B. Johansson, and J. Koll\'ar,
Comp. Mat. Sci. {\bf 18}, 24 (2000); \prb {\bf 29}, 179 (2001).

\bibitem{weinberger90}  P. Weinberger, in {\it Electron scattering theory 
for ordered and disordered matter} (Clarendon Press, Oxford, 1990). 
 
\bibitem{Taylor83}  J.R. Taylor, {\it Scattering Theory: The Quantum Theory 
of Non-relativistic collision} (Robert E. Krieger Pub. Comp, 1983). 

\bibitem{Katsnelson01}  M. I. Katsnelson and A. I. Lichtenstein,
Eur. Phys. J. Phys. B. {\bf 30}, 9 (2002).

\bibitem{berlin}  A.I. Lichtenstein and M. I. Katsnelson, in {\it Band
Ferromagnetsim. Ground State and Finite-Temperature Phenomena}, edited by K.
Barbeschke, M. Donath, and W. Nolting,  Lecture Notes in Physics
(Springer-Verlag, Berlin, 2001), p. 75.

\bibitem{Perdew92} J. P. Perdew and Y. Wang,  Phys. Rev. {\bf 45}, 13244 (1992).

\bibitem{Wu95} R. Wu and A.J. Freeman, \prb {\bf 51}, 17131 (1995).
 
\bibitem{Mirbt97}  S. Mirbt, I. A. Abrikosov, B. Johanson and H. L. Skriver
Phys. Rev. B {\bf 55}, 67 (1997).

\bibitem{Ounadjela}  K. Ounadjela et al, Europhys. Lett. {\bf 15}, 875 
(1991). 
 
\bibitem{Landes}  J. Landes et al, J. Magn. \& Magn. Mater., {\bf 86}, 71 
(1990). 
 
\bibitem{Yafet}  Y. Yafet, J. Appl. Phys. {\bf 61}, 4085 (1987). 
 
\bibitem{Wang}  Y. Wang, P.M. Levy, and J.L. Fry, J. Magn. \& Magn. Mater.  
{\bf 93}, 395 (1991). 
 
\bibitem{Stoeffler}  D. Stoeffler and F. Gautier, Prog. Theor. Phys. Suppl.  
{\bf 101}, 139 (1990). 
 
\bibitem{Levy90} P.M. Levy, S. Zhang and A. Fert, Phys. Rev. Lett.,
{\bf 65}, 1643 (1990).

\bibitem{Itoh95} H. Itoh, J.Inoue and S. Maekawa, Phys. Rev. B,
{\bf 51}, 342 (1995).

\bibitem{Kentzinger03} E. Kentzinger, U. Rucker, B. Topereverg, Th. Bluckel, 
Physica B {\bf 335}, 89 (2003).

\bibitem{Hasegawa88} H.Hasegawa and F. Herman, Phys. Rev. B, {\bf 38}, 4863 (1988);
H.Hasegawa, Phys. Rev. B, {\bf 47} 15080 (1993). 
 
\bibitem{Moriya85} {\it Spin fluctuations in Itinerant Electron Magnetism}, edited by 
T.Moriya (Springer, Berlin, 1985).

\bibitem{Hubbard79} J. Hubbard, Phys. Rev. B, {\bf 19}, 2626 (1979); {\bf 20}, 4583 (1979);
{\bf 23}, 5970 (1983).

\bibitem{phillips71} N. E. Phillips, Crit. Rev. Solid. State Sci.
{\bf 2}, 467 (1971).

\bibitem{fawcett88} E. Fawcett,
Rev. Mod. Phys. {\bf 60}, 209 (1988).
\end{references}
\end{document}